\documentclass[journal,hidelinks]{IEEEtran}

\usepackage{cite}
\usepackage{amsmath,amssymb,amsfonts}
\usepackage{algorithm}
\usepackage{algpseudocode}
\usepackage{graphicx}
\usepackage{textcomp}
\usepackage{url}
\usepackage{comment}
\usepackage{booktabs}
\usepackage{colortbl}
\usepackage{float}
\usepackage{tabularx}
\usepackage{soul}    
\sethlcolor{yellow}  
\usepackage{xcolor}
\usepackage{balance}

\def\BibTeX{{\rm B\kern-.05em{\sc i\kern-.025em b}\kern-.08em
    T\kern-.1667em\lower.7ex\hbox{E}\kern-.125emX}}

\begin{document}

\title{Hybrid Deep Learning Framework for CSI-Based Activity Recognition in Bandwidth-Constrained Wi-Fi Sensing}

\author{
Alison M. Fernandes, Hermes I. Del Monego, Bruno S. Chang, Anelise Munaretto, \\Hélder M. Fontes, and Rui Campos

\thanks{The authors are with Federal University of Technology – Paraná (UTFPR), Graduate Program in Electrical Engineering and Industrial Informatics (CPGEI-CT); Institute of Systems Engineering and Computer Technology, and Science of Portugal (INESC TEC); and Faculty of Engineering of the University of Porto, Portugal.}
\thanks{Corresponding author: Bruno S. Chang (e-mail: bschang@utfpr.edu.br).}

\thanks{This work was supported by the Coordination for the Improvement of Higher Education Personnel (CAPES); Institute for Systems and Computer Engineering, Research and Development of Brazil (INESC P\&D Brazil); and National Council for Scientific and Technological Development (CNPq) under proc. 405940/2022-0.}
}

\pagenumbering{gobble}

\maketitle
\begin{abstract}

This paper presents a novel hybrid deep learning framework designed to enhance the robustness of CSI-based Human Activity Recognition (HAR) within bandwidth-constrained Wi-Fi sensing environments. The core of our proposed methodology is a preliminary Doppler trace extraction stage, implemented to amplify salient motion-related signal features before classification. Subsequently, these enhanced inputs are processed by a hybrid neural architecture, which integrates Inception networks responsible for hierarchical spatial feature extraction and Bidirectional Long Short-Term Memory (BiLSTM) networks that capture temporal dependencies. A Support Vector Machine (SVM) is then utilized as the final classification layer to optimize decision boundaries. The framework's efficacy was systematically validated using a public dataset across 20, 40, and 80 MHz bandwidth configurations. The model yielded accuracies of 89.27\% (20 MHz), 94.13\% (40 MHz), and 95.30\% (80 MHz), respectively. These results confirm a marked superiority over standalone deep learning baselines, especially in the most constrained low-bandwidth scenarios. This study underscores the utility of combining Doppler-based feature engineering with a hybrid learning architecture for reliable HAR in bandwidth-limited wireless sensing applications.
\end{abstract}


\section{Introduction}
\label{sec:introduction}

Wi-Fi sensing has proven in recent years to be a highly effective method for enabling signal-based environmental awareness through Channel State Information (CSI). The ability to capture human activities in a non-intrusive manner, without compromising user privacy, has made it one of the most relevant data acquisition strategies in ambient intelligence applications. CSI reflects how wireless signals propagate through the environment, and its sensitivity to motion makes it suitable for Human Activity Recognition (HAR) using only commodity Wi-Fi devices~\cite{Yang}.

Several recent works ~\cite{Cominelli, Hannan, Liu, Oleg} have explored neural network architectures and signal processing techniques to improve CSI-based recognition systems. However, a significant challenge lies in operating these systems across varying bandwidths. Narrower bandwidths tend to limit the capture of distinctive signal patterns, reducing classification performance. Although it is well known that wider bandwidths generally yield better results, real-world constraints such as legacy hardware, interference, and spectrum regulation often impose bandwidth limitations. This creates a critical need for solutions that maintain high recognition accuracy under constrained bandwidth conditions.

To address this challenge, we propose IBIS, a hybrid deep learning framework for CSI-based activity recognition under bandwidth constraints. The framework uses a Doppler trace extraction stage to enhance motion-sensitive features from CSI phase data. These enhanced representations are then processed by an Inception network for spatial feature extraction and a BiLSTM network for temporal modeling, followed by a SVM for refined classification. The SVM is optimized via Grid Search to determine the best kernel and hyperparameters, effectively improving decision boundaries—particularly in low-bandwidth conditions where conventional deep models tend to underperform \cite{alison}. IBIS is evaluated using bandwidths of 20, 40, and 80 MHz on Scenario S7 from the dataset by Meneghello et al.~\cite{Meneghello}, which presents high interference and environmental variability. The framework achieves recognition accuracies of 89.27\%, 94.13\%, and 95.30\%, respectively, with the most significant relative gain observed at 20 MHz, confirming its robustness in constrained wireless sensing scenarios.

The main contributions of this paper are three-fold: 1) The hybrid deep learning framework (IBIS) that combines Doppler trace extraction from CSI signals with Inception, BiLSTM, and SVM components to enhance recognition accuracy under bandwidth limitations; 2) The systematic evaluation of bandwidth impact (20/40/80 MHz) on HAR accuracy using a challenging real-world dataset; 3) The demonstration that IBIS significantly outperforms conventional deep learning baselines, particularly in low-bandwidth scenarios, highlighting its applicability in real-world, bandwidth-constrained Wi-Fi sensing environments.

This paper is structured as follows. Section \ref{sec:concepts} presents the main concepts related to Wi-Fi sensing, the primary neural networks used, and the mathematical foundations for formulating CSI matrices. Section \ref{sec:experiment} details the experimental setup, including dataset collection and Doppler signal processing. The results are presented in Section \ref{sec:evaluation}, including confusion matrices for different bandwidths. Finally, Section \ref{sec:conclusion} provides the conclusions drawn from the results and suggests directions for future work.

\section{Proposed Hybrid Framework}
\label{sec:concepts}
This section presents the proposed hybrid framework for CSI-based activity recognition. We first review key principles of Wi-Fi sensing and CSI representation, followed by a detailed description of the IBIS architecture, which integrates Inception, BiLSTM, and SVM components to improve recognition performance under bandwidth constraints.

\subsection{Background}

Wi-Fi sensing enables motion detection by leveraging variations in wireless signal propagation, and can be implemented using a range of commodity hardware platforms~\cite{Yang}. Most existing CSI extraction tools operate over relatively narrow bandwidths and support a limited number of subcarriers, which constrains the resolution and, consequently, the accuracy of activity recognition systems. Deployments may involve single or multiple antenna configurations; however, large-scale adoption remains limited by communication overhead and hardware availability. Despite these challenges, Wi-Fi sensing has gained significant traction in the academic community and continues to evolve as a viable technique for non-intrusive human activity monitoring. Its core principle lies in recognizing motion-induced signal patterns within a defined area by capturing CSI between a transmitter and receiver. Fig. \ref{fig:ibis} illustrates this data acquisition process, from wireless transmission to CSI extraction for activity inference.

\begin{figure*}[ht!]
    \begin{center}
    \includegraphics[width =1.0\textwidth]{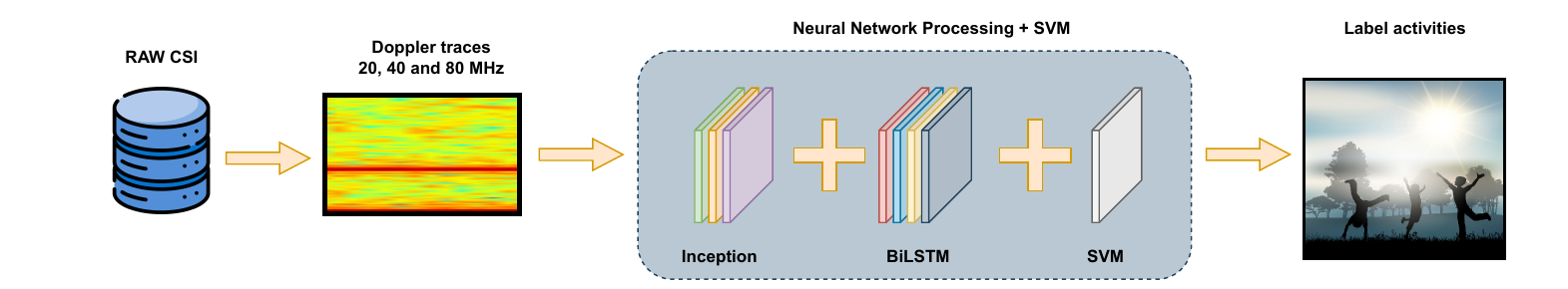}
    \caption{IBIS processing pipeline: CSI signals are collected and transformed via Doppler trace extraction, followed by spatial–temporal feature learning through Inception–BiLSTM networks and final activity classification using an SVM.}
    \label{fig:ibis}
    \end{center}    
\end{figure*}

CSI describes the channel properties of a communication link, reflecting how wireless signals propagate in a physical environment when subject to obstacles, reflection, diffraction, and scattering~\cite{Yang}. In Orthogonal Frequency-Division Multiplexing (OFDM) systems, multiple antennas are typically used at both the transmitter and receiver sides. For each pair of antennas, CSI captures the wireless channel characteristics, including time delay, amplitude attenuation, and phase shift of multipath components on each communication subcarrier.

Mathematically, the channel can be modeled as the Channel Impulse Response (CIR) $h(\tau)$ in the frequency domain, expressed as: 

\begin{equation}
h(\tau) = \sum_{l=1}^{L} \alpha_l e^{j\phi_l} \delta(\tau - \tau_l)
\end{equation}
where $\alpha_l$ and $\phi_l$ represent the amplitude and phase of the $l$-th multipath component, respectively; $\tau_l$ is the time delay; $L$ is the total number of multipath components; and 
$\sigma(\tau)$ denotes the Dirac delta function. In real-world scenarios, the amplitude and phase can be represented as complex numbers, as follows:

\begin{equation}
H_i = |H_i| e^{j \angle H_i}
\end{equation}
where $|H_i|$ and $\angle H_i$ denote the amplitude and phase of the $lth$ subcarrier, respectively. Typical CSI compression for communication purposes aims to estimate this channel matrix for each transmitted packet.

\subsection{IBIS Architecture}

Recent advances in deep learning have led to a variety of architectures being applied to CSI-based HAR, each targeting different feature domains. In particular, Inception modules are known for their ability to perform multiscale spatial feature extraction by applying parallel convolutional layers with varying kernel sizes~\cite{Hornero}. Long Short-Term Memory (LSTM) networks are widely used to capture long-term dependencies in sequential data, enabling the retention of temporal context critical for activity recognition~\cite{Sathyabama}. Support Vector Machines (SVMs) operate by identifying optimal separating hyperplanes in feature space and are particularly effective at maximizing generalization margins in binary and multiclass classification tasks~\cite{Qiyu}.

Other deep learning approaches have also been explored in the literature, including Convolutional Neural Networks (CNNs)~\cite{Chauhan_}, Gated Recurrent Units (GRUs)~\cite{Sathyabama}, Temporal Convolutional Networks (TCNs)~\cite{Niu}, and Transformer-based models, which have gained prominence in generative AI, including BERT and GPT~\cite{Luka}. However, these models often require substantial training data and may struggle to generalize under bandwidth-constrained sensing scenarios.

To address this, we propose IBIS, a hybrid deep learning architecture that integrates an Inception module and a Bidirectional LSTM (BiLSTM) with a post-processing SVM classifier. The objective is to enhance recognition robustness when operating with CSI data derived from limited bandwidth configurations.

As part of its pipeline, IBIS includes a \textbf{Doppler trace extraction} stage that transforms the phase component of the CSI into time–frequency representations. This enhances motion-sensitive features and provides richer inputs to the deep learning model. The \textbf{Inception module} then performs multiscale spatial analysis by applying 1D convolutions in parallel across the Doppler-enhanced input, enabling the extraction of local and global frequency-domain patterns associated with human motion. The \textbf{BiLSTM layer} models the temporal evolution of CSI sequences in both forward and backward directions, capturing dependencies across time that are critical for distinguishing between dynamic activities such as Walking, Running, and Jumping. Finally, the \textbf{SVM post-processing} stage refines classification boundaries using the probabilistic output of the BiLSTM. Kernel selection and hyperparameter tuning are performed via Grid Search, selecting from polynomial, Radial Basis Function (RBF), or sigmoid kernels to maximize generalization performance across different bandwidths.

The complete IBIS pipeline is depicted in Fig. \ref{fig:ibis}, which illustrates the flow from raw CSI data acquisition and Doppler trace extraction, through the hybrid neural network processing, to the final activity classification. The model operates on Doppler-enhanced CSI matrices where only the phase information is preserved, given its superior sensitivity to motion-induced variations. Amplitude information is discarded to improve robustness against environmental noise and hardware variability. By combining multiscale spatial learning, bidirectional temporal modeling, and decision boundary optimization, IBIS is designed to deliver stable and accurate performance even in challenging low-bandwidth sensing scenarios.

\begin{figure}[ht!]
    \begin{center}
    \includegraphics[width =0.45\textwidth]{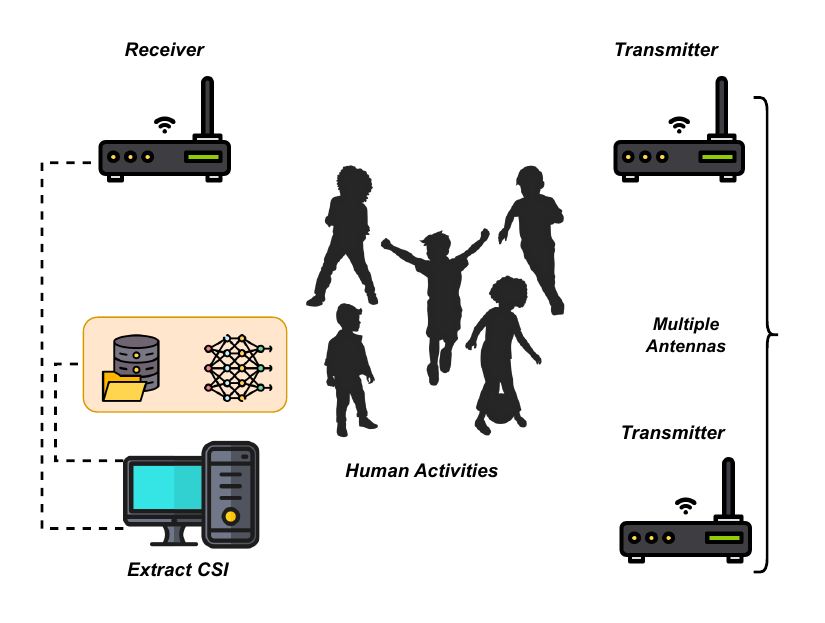}
    \caption{CSI acquisition setup: wireless signals are transmitted and received via multiple antennas, enabling CSI extraction for subsequent Doppler-based processing and activity recognition.}
    \label{fig:ibis}
    \end{center}    
\end{figure}

\section{Experimental Setup}
\label{sec:experiment}

This section describes the experimental procedure used to evaluate the IBIS framework. It includes the dataset characteristics, the Doppler feature extraction process, and the overall training and testing pipeline used across different bandwidth configurations.

\subsection{Dataset Collection}

The dataset used in this work was originally published by Meneghello et al.~\cite{Meneghello}, and includes seven distinct activity scenarios (S1–S7) recorded in indoor environments with varying levels of complexity. For evaluation, Scenario S7 was selected, as it represents the most challenging environment due to its high density of reflective surfaces and obstacles, which significantly impact CSI quality and classification performance. An additional advantage of this dataset is the ability to simulate different bandwidths—20, 40, and 80 MHz—by selectively aggregating subcarriers in the spectrum.

\subsection{Doppler Traces Generation}

\begin{figure*}[ht!]
    \centering
    
    \begin{minipage}[b]{1.0\textwidth}
        \centering
        \includegraphics[width=\textwidth]{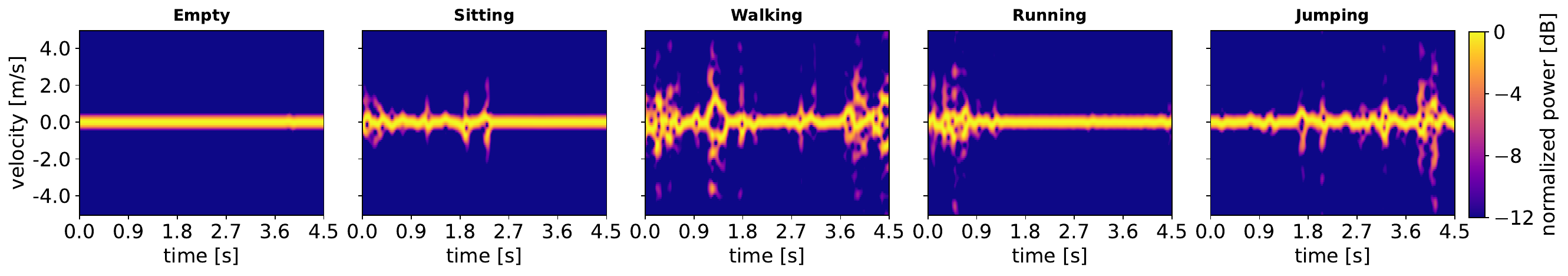}
        \text{(a)}
        \label{fig:conf_doppler_a}
    \end{minipage}

    \begin{minipage}[b]{1.0\textwidth}
        \centering
        \includegraphics[width=\textwidth]{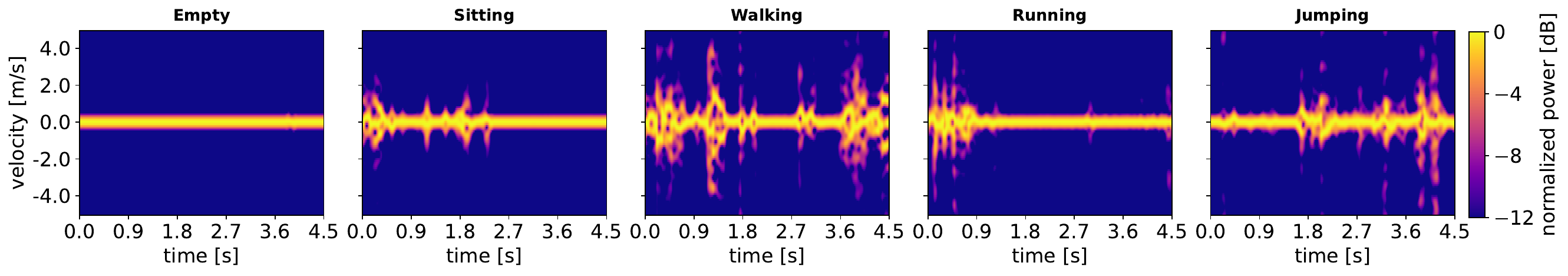}
        \text{(b)}
        \label{fig:conf_doppler_b}
    \end{minipage}

    \begin{minipage}[b]{1.0\textwidth}
        \centering
        \includegraphics[width=\textwidth]{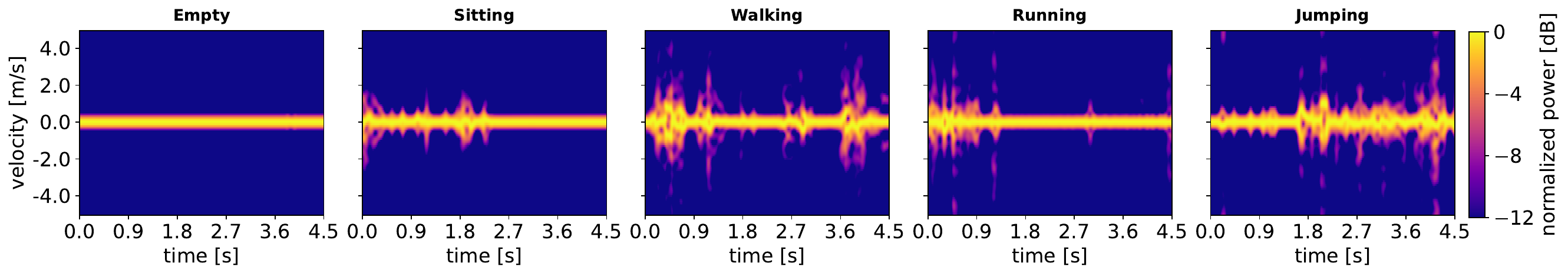}
        \text{(c)}
        \label{fig:conf_doppler_c}
    \end{minipage}
    
    \caption{Normalized Doppler spectrograms for five activities—\emph{Empty}, \emph{Sitting}, \emph{Walking}, \emph{Running}, and \emph{Jumping}—captured at three bandwidth configurations: (a) 20 MHz, (b) 40 MHz, and (c) 80 MHz. Higher motion levels introduce greater Doppler variation, increasing classification complexity.}
    \label{fig:conf_doppler}
\end{figure*}

To evaluate the impact of bandwidth on recognition accuracy, the experiments were structured in three stages corresponding to 20, 40, and 80 MHz configurations. For each configuration, Doppler traces were generated for five activities: \emph{Empty}, \emph{Sitting}, \emph{Walking}, \emph{Running}, and \emph{Jumping}. Each Doppler spectrogram spans a 4.5-second window with a velocity range of –4 to 4 m/s. As shown in Fig. \ref{fig:conf_doppler}, high-motion activities such as \emph{Walking}, \emph{Running}, and \emph{Jumping} exhibit greater Doppler variation, increasing class overlap and making classification more challenging. These traces were computed from the phase component of CSI signals, which is more sensitive to motion dynamics.

In total, 234 samples were collected using an 80 MHz channel compliant with the IEEE 802.11ac standard and extracted via the \emph{Nexmon} tool. Subcarrier gaps were preserved to ensure spectral orthogonality and reduce adjacent-channel interference. These include guard bands, pilot carriers, and other elements needed to avoid signal distortion due to filtering and hardware imperfections.

\subsection{Experimental Pipeline}
The experimental pipeline follows three main stages: preprocessing, training, and testing.

In the preprocessing stage, noise and offset components in the CSI phase data were removed using filtering and regularization techniques, including Least Absolute Shrinkage and Selection Operator (LASSO). Only the phase information was retained, as the amplitude component is more susceptible to noise and environmental inconsistencies.

After Doppler trace extraction, the data were organized into training and testing sets. Each activity class was segmented into blocks, and data were augmented across the three bandwidth configurations. IBIS was then trained using this dataset to extract spatial and temporal features. Output probabilities from the BiLSTM were used to train a SVM classifier. Then, a Grid Search was conducted separately for each bandwidth configuration to identify the optimal SVM kernel (RBF, sigmoid, or polynomial) and hyperparameters $C$ and $\gamma$. Each experiment was repeated ten times, and final performance metrics were averaged across all runs.

In the final stage, inference was performed using the most challenging environment (Scenario S7). The trained SVM was applied to the BiLSTM probability outputs to refine the classification decisions on unseen data. To further enhance performance, an antenna merge strategy was used, aggregating predictions from different antennas based on a majority voting criterion.

\section{Experimental Evaluation}
\label{sec:evaluation}

This section presents the evaluation results of the IBIS framework for three different bandwidth configurations: 20, 40, and 80 MHz. The evaluation compares IBIS to a baseline Inception-only model, considering accuracy, the influence of antennas, and the individual contributions of the Doppler feature extraction and SVM classifier components. All experiments were conducted using Scenario S7 from the dataset proposed by Meneghello et al.~\cite{Meneghello}, which includes environmental complexity and high signal interference.

\subsection{Impact of Bandwidth on Classification Accuracy}
\begin{figure}[ht!]
    \begin{center}
    \includegraphics[width =0.5\textwidth]{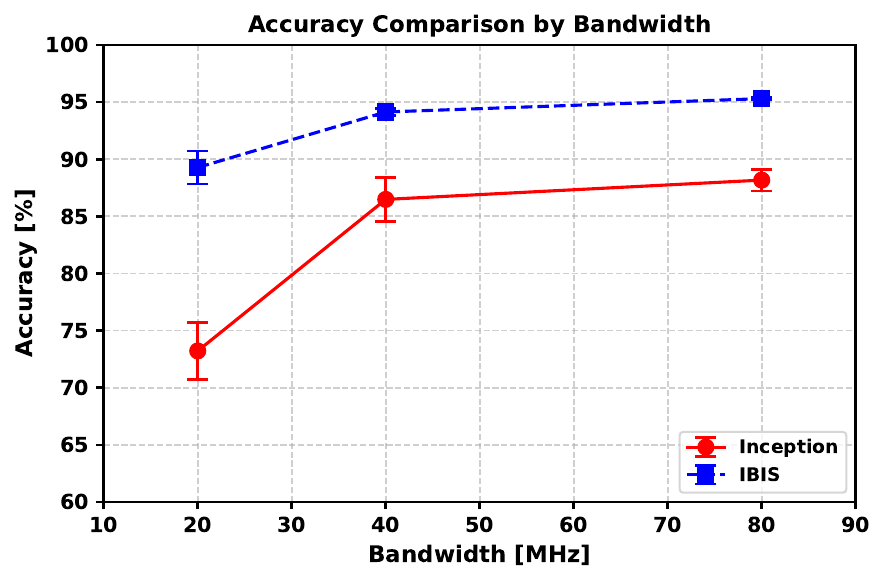}
    \caption{Classification accuracy of Inception and IBIS across bandwidths of 20, 40, and 80 MHz. IBIS consistently outperforms the baseline, with the largest gain at 20 MHz.}
    \label{fig:comp_in_ibis}
    \end{center}    
\end{figure}

Fig. \ref{fig:comp_in_ibis} compares the average accuracy of IBIS and a baseline Inception-only model across the evaluated bandwidths. 

At 20 MHz, the baseline model reaches an accuracy of 73.22\%, while IBIS achieves 89.27\%, a gain of 16.07 percentage points. This substantial improvement is attributed to the combined benefits of Doppler trace extraction, temporal modeling with BiLSTM, and SVM-based decision refinement. Interestingly, the best performance at this bandwidth was achieved using a polynomial kernel, which outperformed the more commonly used RBF kernel due to its better separation of overlapping activity classes in the low-dimensional feature space.

At 40 MHz, IBIS attains an accuracy of 94.13\%, improving upon the 86.48\% achieved by the baseline—an advantage of 7.65 percentage points. In this setting, the RBF kernel performed best, enabling the SVM to refine the BiLSTM output more effectively.

At 80 MHz, IBIS achieves 95.30\%, slightly higher than the 40 MHz result, indicating that most of the relevant spatial–temporal information is already well captured at 40 MHz. The performance improvement at 80 MHz is mainly due to more effective post-processing via SVM. Table \ref{tab:tradeoff_gamma} details the hyperparameter tuning process, showing that the best results were obtained with hyperparameters $C$ and $\gamma$ on accuracy, showing that $C=1$ and $\gamma=1$ yield optimal performance.

The observed improvement in the IBIS framework can be attributed to two main factors: the BiLSTM module effectively captures temporal dependencies in the Doppler traces, enhancing the Inception network’s ability to learn more complex temporal-spatial features; and the SVM classifier further refines the decision boundaries, providing optimal separation between classes and thereby improving overall classification performance.

\begin{table}[hbt!]
\caption{SVM classification accuracy across different hyperparameter settings ($C$ and $\gamma$) for each bandwidth. Optimal performance is obtained with $C=1$ and $\gamma=1$}
\label{tab:tradeoff_gamma}
\begin{center}
\begin{tabular}{p{1.8cm}p{1.8cm}p{1.8cm}p{1.8cm}}
\toprule
$C$ & $\gamma = 0.01$ (\%) & $\gamma = 0.1$ (\%) & $\gamma = 1$ (\%) \\
\midrule
0.01 & 88.49 & 80.52 & 91.41 \\
0.1  & 90.69 & 72.90 & 91.80 \\
1    & 83.46 & 80.06 & 95.54 \\
\bottomrule
\end{tabular}
\end{center}
\footnotesize{Note: $C$ – Regularization, $\gamma$ – Kernel influence}
\end{table}

\subsection{Comparative Analysis}
Table \ref{tab:metric_bw} summarizes the average performance metrics of each neural network model across the three evaluated bandwidth configurations. As shown, for the Inception network, the accuracy decreases as the bandwidth is reduced. The drop from 80 MHz to 40 MHz is minor—approximately 1.69\%—whereas the decrease from 40 MHz to 20 MHz is more pronounced, at about 13.26\%. This trend is consistent with the observations of Cominelli~\cite{Cominelli}, who analyzed the impact of bandwidth on neural network performance. According to that study, the information relevant for activity pattern recognition tends to be concentrated in specific sub-bands depending on spatial position. Wider bandwidths provide richer spectral information, allowing the neural network to extract more stable and discriminative features. Conversely, narrower bandwidths may lose crucial signal details, leading to higher susceptibility to overfitting during training. Hence, while increased bandwidth does not always guarantee better performance, broader channels generally improve recognition accuracy by offering a more informative feature space. These findings reinforce the ongoing efforts in the research community to enhance CSI-based sensing accuracy under narrowband constraints.

Another important observation is that the IBIS model at 20 MHz achieves an accuracy of 89.27\%, which surpasses the 88.17\% obtained by the Inception network at 80 MHz. This demonstrates the robustness and effectiveness of the proposed hybrid framework. IBIS maintains high recognition performance even with reduced spectral resolution, benefiting from its Doppler-based feature enhancement, spatial–temporal modeling, and SVM-based post-processing.

Overall, the results in Table \ref{tab:metric_bw} confirm that IBIS consistently outperforms the baseline model across all evaluated metrics—accuracy, precision, recall, and F1 score—with the largest relative improvement observed in the 20 MHz configuration.

\begin{table}[hbt!]
\caption{Average accuracy, precision, recall, and F1 score for Inception and IBIS across bandwidths. IBIS outperforms the baseline in all metrics, with the largest relative gains at 20 MHz.}
\label{tab:metric_bw}
\centering
\begin{minipage}{0.5\textwidth} 
\scriptsize
\setlength{\tabcolsep}{4pt}
\renewcommand{\arraystretch}{0.9}
\begin{tabular}{p{2.5cm} c c c c c c}
\toprule
Bandwidth & \multicolumn{2}{c}{20 MHz} & \multicolumn{2}{c}{40 MHz} & \multicolumn{2}{c}{80 MHz} \\
\midrule
Models & Inception & IBIS & Inception & IBIS & Inception & IBIS \\
\midrule
Accuracy (\%) & 73.22 & 89.27 & 86.48 & 94.13 & 88.17 & 95.13 \\
Precision (\%) & 80.57 & 90.18 & 88.80 & 94.84 & 91.47 & 95.57 \\
Recall (\%) & 73.58 & 89.07 & 86.29 & 94.04 & 87.87 & 95.30 \\
F1 Score (\%) & 70.65 & 89.23 & 86.36 & 93.83 & 88.26 & 95.21 \\
\bottomrule
\end{tabular}
\end{minipage}
\end{table}

\subsection{Influence of Antenna Diversity}

\begin{figure}[ht!]
    \begin{center}
    \includegraphics[width =0.5\textwidth]{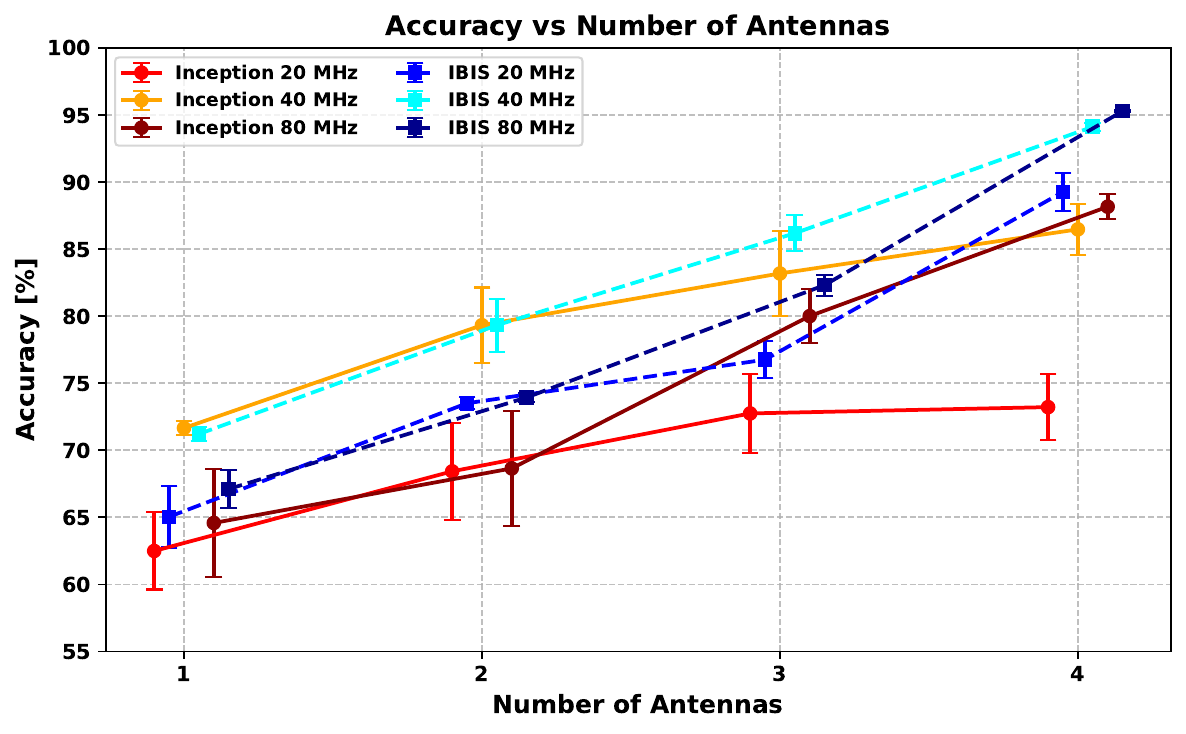}
    \caption{Average classification accuracy of Inception and IBIS as a function of the number of antennas. IBIS benefits significantly from antenna diversity, achieving its highest performance with four antennas.}
    \label{fig:ant}
    \end{center}    
\end{figure}

Fig. \ref{fig:ant} illustrates the effect of increasing antenna count on classification accuracy. IBIS shows progressive performance gains with each additional antenna, achieving its peak accuracy of 95.30\% using four antennas. In contrast, single-antenna configurations yield lower performance in both models, confirming that spatial diversity is a critical factor in CSI-based activity recognition systems.

\subsection{Effect of Doppler Traces and SVM Removal}

\begin{figure}[ht!]
    \begin{center}
    \includegraphics[width =0.5\textwidth]{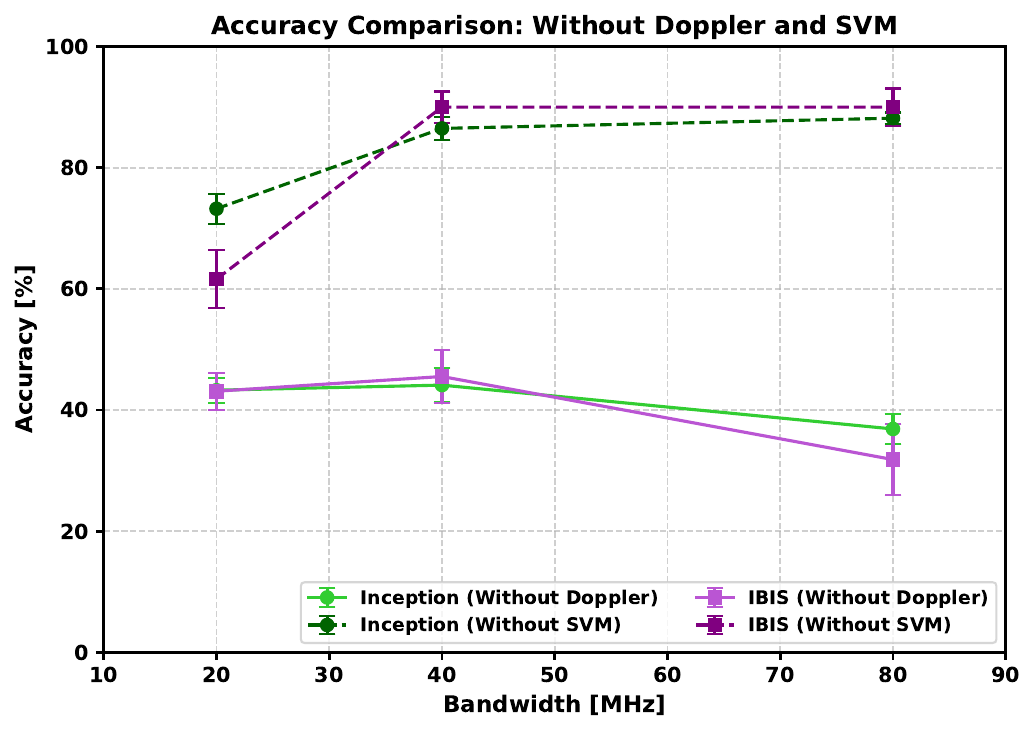}
    \caption{Ablation analysis of IBIS components. Removing Doppler trace extraction or SVM post-processing significantly degrades performance, especially under low-bandwidth conditions.}
    \label{fig:comp_svm_doppler}
    \end{center}    
\end{figure}

Fig. \ref{fig:comp_svm_doppler} presents the results of two ablation experiments designed to isolate the contribution of Doppler trace extraction and SVM-based post-processing in the IBIS pipeline.

In the first scenario, Doppler traces were removed from the input data. Only sanitized raw CSI signals were used, omitting any frequency–time transformation. Under these conditions, at 20 MHz, both the IBIS and Inception models achieve very similar accuracy levels, with no significant distinction between them. This suggests that without Doppler-based enhancement, the models rely mainly on majority class prediction, resulting in correct classification of only the most dominant activities—approximately 40\% of the dataset. Similar patterns are observed for the 40 MHz and 80 MHz configurations, indicating that the removal of Doppler information severely degrades the model’s discriminative capacity. These findings confirm that Doppler trace extraction plays a crucial role in enhancing activity-specific features and enabling meaningful classification, particularly in bandwidth-limited scenarios.

In the second scenario, the SVM component was removed from the pipeline, leaving only the Inception–BiLSTM network. As shown in Fig. \ref{fig:comp_svm_doppler}, the resulting model achieved 61.60\% accuracy at 20 MHz, which is lower than the 73.22\% obtained using the Inception-only baseline, and significantly below the 89.27\% reached by the complete IBIS framework (which includes SVM with a polynomial kernel). For 40 MHz and 80 MHz, the accuracy gap was smaller—just below 90\%—suggesting that the SVM contributes most when the network lacks strong discriminative features, as is typically the case in narrowband conditions.

These results reinforce the role of each IBIS component: the Inception module extracts global spatial patterns; the BiLSTM captures temporal dependencies derived from Doppler variations; and the SVM refines decision boundaries to maximize class separation. The synergy among these elements is essential to fully exploit the potential of the IBIS framework.

\section{Conclusion}
\label{sec:conclusion}

This paper introduced IBIS, a novel hybrid deep learning framework engineered to mitigate performance degradation in CSI-based human activity recognition under bandwidth-constrained conditions. The proposed architecture synergistically integrates Doppler trace extraction with an Inception-BiLSTM hybrid network and a SVM classifier. This design enhances both spatial-temporal feature representation and final classification robustness. Experimental validation across 20, 40, and 80 MHz configurations demonstrated that IBIS consistently and significantly outperforms a standard deep learning baseline. Crucially, the framework achieved 89.27\% accuracy at 20 MHz, a result that surpasses the baseline's performance even at its optimal 80 MHz configuration. This analysis confirms the indispensable roles of both Doppler-based feature engineering and SVM-based refinement in achieving high recognition accuracy in challenging, low-bandwidth scenarios.

Future work will explore the generalization of the IBIS framework to multi-person activity scenarios and more dynamic environments. We also plan to investigate the integration of transformer-based architectures and self-supervised learning strategies to reduce dependency on labeled data, as well as on-device deployment strategies to optimize inference efficiency in real-time applications.

\balance

\bibliographystyle{IEEEtran} 
\bibliography{references}

\end{document}